\documentclass{article}
\usepackage{ismir,amsmath,cite,url,amssymb}
\usepackage{graphicx}
\usepackage{color}
\usepackage{enumitem}

\title{Using musical relationships between chord labels in Automatic Chord Extraction tasks}

 \multauthor
 {Tristan Carsault$^1$ \hspace{1cm} J{\'e}r{\^o}me Nika$^{2,1}$ \hspace{1cm} Philippe Esling$^1$} { 
 $^1$Ircam, CNRS, Sorbonne Universit{\'e}, UMR 9912 STMS,
 $^2$L3i Lab, University of La Rochelle\\
 {\tt \{carsault, jnika, esling\}@ircam.fr}
 }
 \def\authorname{Tristan Carsault, J{\'e}r{\^o}me Nika, Philippe Esling}

\begin{document}

\maketitle
\begin{abstract}

Recent research on Automatic Chord Extraction (ACE) has focused on the improvement of models based on machine learning. However, most models still fail to take into account the prior knowledge underlying the labeling alphabets (chord labels). Furthermore, recent works have shown that ACE performances have reached a glass ceiling. Therefore, this prompts the need to focus on other aspects of the task, such as the introduction of musical knowledge in the representation, the improvement of the models towards more complex chord alphabets and the development of more adapted evaluation methods. 

In this paper, we propose to exploit specific properties and relationships between chord labels in order to improve the learning of statistical ACE models. Hence, we analyze the interdependence of the representations of chords and their associated distances, the precision of the chord alphabets, and the impact of performing alphabet reduction before or after training the model. Furthermore, we propose new training losses based on musical theory. We show that these improve the results of ACE systems based on Convolutional Neural Networks. 
By analyzing our results, we uncover a set of related insights on ACE tasks based on statistical models, and also formalize the musical meaning of some classification errors.

%
%
%



\end{abstract}
\section{Introduction}\label{sec:introduction}


Automatic Chord Extraction (ACE) is a topic that has been widely studied by the Music Information Retrieval (MIR) community over the past years. However, recent results seem to indicate that the rate of improvement of ACE performances has diminished over the past years \cite{mcvicar2014automatic}. 

Recently, a part of the MIR community pointed out the need to rethink the experimental methodologies. Indeed, current evaluation methods do not account for the intrinsic relationships between different chords \cite{humphrey2015four}. Our work is built on these questions and is aimed to give some insights on the impact of introducing musical relationships between chord labels in the development of ACE methods. 

Most ACE systems are built on the idea of extracting features from the raw audio signal and then using these features to construct a chord classifier \cite{cho2010exploring}. The two major families of approaches that can be found in previous research are \emph{rule-based} and \emph{statistical} models. On one hand, the rule-based models rely on music-theoretic rules to extract information from the precomputed features.
Although this approach is theoretically sound, it usually remains brittle to perturbations in the spectral distributions from which the features were extracted.
On the other hand, statistical models rely on the optimization of a loss function over an annotated dataset. However, the generalization capabilities of these models are highly correlated to the size and completeness of their training set. Furthermore, most training methods see musical chords as independent labels and do not take into account the inherent relations between chords.

In this paper, we aim to target this gap by introducing musical information directly in the training process of statistical models. To do so, we propose to use prior knowledge underlying the labeling alphabets in order to account for the inherent relationships between chords directly inside the loss function of learning methods. Due to the complexity of the ACE task and the wealth of models available, we choose to rely on a single Convolutional Neural Network (CNN) architecture,
which provides the current best results in ACE \cite{mcfee2017structured}. First, we study the impact of chord alphabets and their relationships by introducing a specific \emph{hierarchy} of alphabets. We show that some of the reductions proposed by previous researches might be inadequate for learning algorithms. We also show that relying on more finely defined and extensive alphabets allows to grasp more interesting insights on the errors made by ACE systems, even though their accuracy is only marginally better or worse.
Then, we introduce two novel chord distances based on musical relationships found in the \emph{Tonnetz-space} or directly between chord components through their categorical differences. These distances can be used to define novel loss functions for learning algorithms. We show that these new loss functions improve ACE results with CNNs.
Finally, we perform an extensive analysis of our approach and extract insights on the methodology required for ACE. To do so, we develop a specifically-tailored analyzer that focuses on the functional relations between chords to distinguish \emph{strong} and \emph{weak} errors. This analyzer is intended to be used for future ACE research to develop a finer understanding on the reasons behind the success or failure of ACE systems.



\section{Related Works}

Automatic Chord Extraction (ACE) is defined as the task of labeling each segment of an audio signal using an alphabet of musical chords. In this task, chords are seen as the concomitant or successive combination of different notes played by one or many instruments.

\subsection{Considerations on the ACE task}
Whereas most MIR tasks have benefited continuously from the recent advances in deep learning, the ACE field seems to have reached a glass ceiling. In 2015, Humphrey and Bello \cite{humphrey2015four} highlighted the need to rethink the whole ACE methodology by giving four insights on the task. 

First, several songs from the reference annotated chord datasets (Isophonics, RWC-Pop, McGill Billboard) are not always tuned to 440Hz and may vary up to a quarter-tone. This leads to multiple misclassifications on the concomitant semi-tones. Moreover, chord labels are not always well suited to describe every song in these datasets.

Second, the chord labels are related and some subsets of those have hierarchical organizations.
Therefore, the one-to-K assessment where all errors are equivalently weighted appears widely incorrect. For instance, the misclassification of a \textit{C:Maj} as a \textit{A:min} or \textit{C\#:Maj}, will be considered equivalently wrong. However, \textit{C:Maj} and \textit{A:min} share two pitches in common whereas \textit{C:Maj} and \textit{C\#:Maj} have totally different pitch vectors.

Third, the very definition of the ACE task is also not entirely clear. Indeed, there is a frequent confusion between two different tasks. First, 
the literal \textit{recognition} of a local audio segment using a chord label and its precise extensions, and, second,
%
the \textit{transcription of an underlying harmony}, taking into account the functional aspect of the chords and the long-term structure of the song. 
%
%
Finally, the labeling process involves the subjectivity of the annotators. For instance, even for expert annotators, it is hard to agree on possible chord inversions.

Therefore, this prompts the need to focus on other aspects such as the introduction of musical knowledge in the representation of chords, the improvement of the models towards more complex chord alphabets and the development of more adapted evaluation methods.

\subsection{Workflow of ACE systems}

Due to the complexity of the task, ACE systems are usually divided into four main modules performing \emph{feature extraction}, \emph{pre-filtering}, \emph{pattern matching} and \emph{post-filtering} \cite{cho2010exploring}.


First, the \emph{pre-filtering} usually applies low-pass filters or harmonic-percussive source separation methods on the raw signal \cite{zhou2015chord, jiang2017extended}. This optional step allows to remove noise or other percussive information that are irrelevant for the chord extraction task. Then, the audio signal is transformed into a time-frequency representation such as the Short-Time Fourier Transform (STFT) or the Constant-Q Transform (CQT) that provides a logarithmically-scaled frequencies. These representations are sometimes summarized in a pitch class vector called \emph{chromagram}. Then, successive time frames of the spectral transform are averaged in context windows. This allows to smooth the extracted features and account for the fact that chords are longer-scale events. It has been shown that this could be done efficiently by feeding STFT context windows to a CNN in order to obtain a clean chromagram \cite{korzeniowski2016feature}. 

Then, these extracted features are classified by relying on either a rule-based chord template system or a statistical model. Rule-based methods give fast results and a decent level of accuracy \cite{oudre2009template}. With these methods, the extracted features are classified using a fixed dictionary of chord profiles \cite{cannam2015mirex} or a collection of decision trees \cite{jiang2017extended}. However, these methods are usually brittle to perturbations in the input spectral distribution and do not generalize well.

Statistical models aim to extract the relations between precomputed features and chord labels based on a training dataset in which each temporal frame is associated to a label. The optimization of this model is then performed by using gradient descent algorithms to find an adequate configuration of its parameters. Several probabilistic models have obtained good performances in ACE, such as multivariate Gaussian Mixture Model \cite{cho2014improved} and convolutional \cite{korzeniowski2016fully, humphrey2012rethinking} or recurrent \cite{wu2017mirex,boulanger2013audio} Neural Networks.


Finally, \emph{post-filtering} is applied to smooth out the classified time frames. This is usually based on a study of the transition probabilities between chords by a Hidden Markov Model (HMM) optimized with the Viterbi algorithm \cite{lou1995implementing} or with Conditional Random Fields \cite{lafferty2001conditional}.




\subsection{Convolutional Neural Network}

A Convolutional Neural Network (CNN) is a statistical model composed of layers of artificial neurons that transform the input by repeatedly applying convolution and pooling operations.  A convolutional layer is characterized by a set of convolution kernels that are applied in parallel to the inputs to produce a set of output \textit{feature maps}.
The convolution kernels are defined as three-dimensional tensors $h \in \mathbb{R}^{M \times U \times V}$ where $M$ is the number of kernels, $U$ is the height and $V$ the width of each kernel.
If we note the input as matrix $X$, then the output feature maps are defined by $Y = X \ast h_m $ for every kernels, where $\ast$ is a 2D discrete convolution operation

\begin{equation}
\label{conv}
(A \ast B)_{i,j} = \sum\limits_{r=0}^{(T-1)} \sum\limits_{s=0}^{(F-1)} A_{r,s} B_{i-r, j-s}
\end{equation}

for  $A \in \mathbb{R}^{T \times F}$ and $B \in \mathbb{R}^{U \times V}$  with $0 \leq i \leq T + U - 1$ and $0 \leq j \leq F + V -1$. \\


As this convolutional layer significantly increases the dimensionality of the input data, a pooling layer is used to reduce the size of the feature maps. The pooling operation reduces the maps by computing local mean, maximum or average of sliding context windows across the maps.
Therefore, the overall structure of a CNN usually consists in alternating convolution, activation and pooling layers. Finally, in order to perform classification, this architecture is typically followed by one or many fully-connected layers. Thus, the last layer produces a probability vector of the same size as the chord alphabet. As we will rely on the architecture defined by \cite{humphrey2012rethinking}, we redirect interested readers to this paper for more information.




\section{Our proposal}

\subsection{Definition of alphabets}\label{sec:alph_red}

\begin{figure}
\centering
\includegraphics[width=0.47\textwidth]{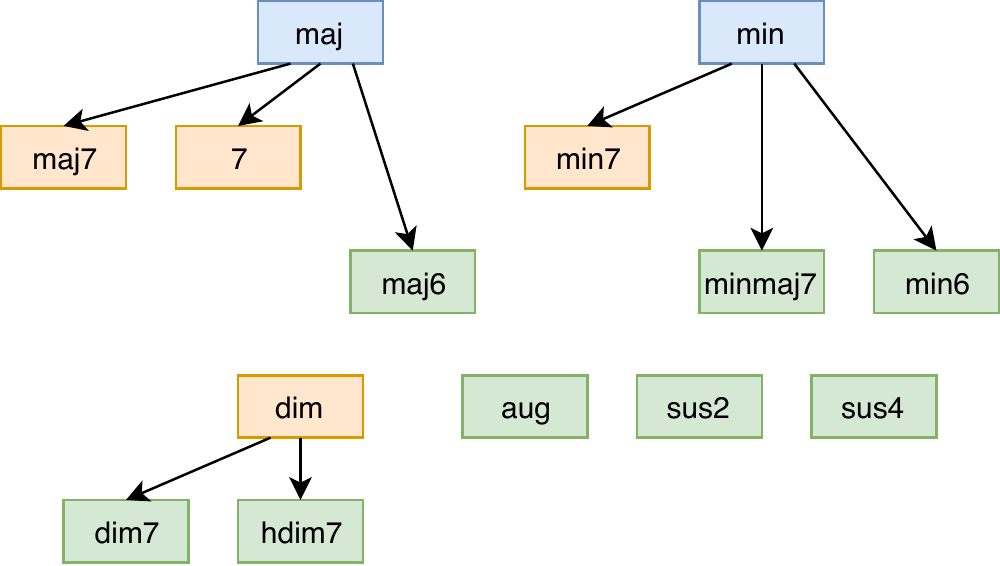}
\caption{\label{fig:alph} Hierarchy of the chord alphabets (blue: $A_0$, orange: $A_1$, green: $A_2$) }
\end{figure}

Chord annotations from reference datasets are very precise and include extra notes (in parenthesis) and basses (after the slash) \cite{harte2005symbolic}. With this notation, we would obtain over a thousand chord classes with very sparse distributions. However, we do not use these extra notes and bass in our classification. Therefore, we can remove this information
\begin{equation}
F:maj7(11)/3 \rightarrow F:maj7
\end{equation}
Even with this reduction, the number of chord qualities (eg. \textit{maj7, min, dim}) is extensive and we usually do not aim for such a degree of precision. Thus, we propose three alphabets named $A_0$, $A_1$ and $A_2$ with a controlled number of chord qualities. The level of precision of the three alphabets increases gradually (see Figure~\ref{fig:alph}). In order to reduce the number of chord qualities, each one is mapped to a parent class when it exists, otherwise to the \textit{no-chord} class $N$.

The first alphabet $A_0$ contains all the major and minor chords, which defines a total of 25 classes

\begin{equation}
A_0 = \{ N \} \cup  \{ P\times maj,min \}
\end{equation}

where $P$ represents the 12 pitch classes.


Here, we consider the interest of working with chord alphabets larger than $A_0$. Therefore, we propose an alphabet containing all chords present in the harmonization of the major scale (usual notation of harmony in jazz music).
This corresponds to the orange chord qualities and their parents in Figure~\ref{fig:alph}. 
The chord qualities without heritage are included in the no-chord class $N$, leading to 73 classes

\begin{equation}
A_1 = \{ N \} \cup  \{ P\times maj, min, dim, maj7, min7, 7 \}
\end{equation}


Finally, the alphabet $A_2$ is inspired from the large vocabulary alphabet proposed by \cite{mcfee2017structured}. This most complete chord alphabet contains 14 chord qualities and 169 classes

\begin{equation}
\begin{split}
A_2 = \{ N \} \cup  \{ P\times maj, min, dim, aug, maj6, min6,\\maj7, minmaj7, min7, 7, dim7, hdim7, sus2, sus4 \}
\end{split}
\end{equation}


\subsection{Definition of chord distances}

In most CNN approaches, the model does not take into account the nature of each class when computing their differences. Therefore, this distance which we called categorical distance $D_0$ is the binary indicator

\begin{equation}
D_0(chord_1,chord_2) = \left\{ 
\begin{array}{l l}
  0 & \quad \text{if } chord_1=chord_2\\
  1 & \quad \text{if } chord_1\ne chord_2\\ \end{array} \right.
\end{equation}

However, we want here to include the relationships between chords directly in our model. For instance, a \textit{C:maj7} is closer to an \textit{A:min7} than a \textit{C\#:maj7}. Therefore, we introduce more refined distances that can be used to define the loss function for learning.\\
Here, we introduce two novel distances that rely on the representation of chords in an harmonic space or in a pitch space to provide a finer description of the chord labels. However, any other distance that measure similarities between chords could be studied \cite{madjiheurem2016chord2vec,huang2016chordripple}.




\subsubsection{Tonnetz distance}

A \emph{Tonnetz-space} is a geometric representation of the tonal space based on harmonic relationships between chords.
We chose a Tonnetz-space generated by three transformations 
of the major and minor triads \cite{cohn1997neo} 
changing only one of the three notes of the chords: 
%
the \textit{relative transformation} (transforms a chord into his relative major / minor), the
\textit{parallel transformation} (same root but major instead of minor or conversely), the
\textit{leading-tone exchange} (in a major chord the root moves down by a semitone, in a minor chord the fifth moves up by a semitone).
%
Representing chords in this space has already shown promising results for classification on the $A_0$ alphabet \cite{humphrey2012learning}.

We define the cost of a path between two chords as the sum of the succesive transformations.
Each transformation is associated to the same cost.
Furthermore, an extra cost is added if the chords have been reduced beforehand in order to fit the alphabet $A_0$. 
Then, our distance $D_1$ is:

\begin{equation}
D_1(chord_1,chord_2) = min(C) 
\end{equation}

with $C$ the set of all possible path costs from $chord_1$ to $chord_2$ using a combination of the three transformations.

\subsubsection{Euclidean distance on pitch class vectors}

In some works, pitch class vectors are used as an intermediate representation for ACE tasks \cite{lee2006automatic}. Here, we use these pitch class profiles to calculate the distances between chords according to their harmonic content. 

Each chord from the dictionary is associated to a 12-dimensional binary pitch vector with 1 if the pitch is present in the chord and 0 otherwise (for instance \textit{C:maj7} becomes $(1,0,0,0,1,0,0,1,0,0,0,1)$). 
The distance between two chords is defined as the Euclidean distance between the two binary pitch vectors.

\begin{equation}
D_2(chord_1,chord_2) = \sqrt{\sum_{i=0}^{11}(chord_{1}^i - chord_{2}^i)^2} 
\end{equation}

Hence, this distance allows to account for the number of pitches that are shared by two chords.


The $D_0$, $D_1$ or $D_2$ distance is used to define the loss function for training the CNN classification model.





\subsection{Introducing the relations between chords}\label{subsec:intr}

To train the model with our distances, we first reduce the original labels from the Isophonics dataset\footnote{\url{http://isophonics.net/content/reference-annotations-beatles}} so that they fit one of our three alphabets $A_0$, $A_1$, $A_2$. Then, we denote $y_{true}$ as the one-hot vector where each bin corresponds to a chord label in the chosen alphabet $A_i$. The output of the model, noted $y_{pred}$, is a vector of probabilities over all the chords in a given alphabet $A_i$.
In the case of $D_0$, we train the model with a loss function that simply compares $y_{pred}$ to the original label $y_{true}$. However, for our proposed distances ($D_1$ and $D_2$), we introduce a similarity matrix $M$ that associates each couple of chords to a similarity ratio. 






\begin{equation}
M_{i,j} = \frac{1}{D_k(chord_i,chord_j) + K}
\end{equation}

K is an arbitrary constant to avoid division by zero. The matrix $M$ is symmetric and we normalize it with its maximum value to obtain $\bar{M}$. Afterwards, we define a new $\bar{y_{true}}$ which is the matrix multiplication of the old $y_{true}$ and the normalized matrix $\bar{M}$.

\begin{equation}
\bar{y_{true}} = y_{true}\bar{M}
\end{equation}


Finally, the loss function for $D_1$ and $D_2$ is defined by a comparison between this new ground truth $\bar{y_{true}}$ and the output $y_{pred}$. Hence, this loss function can be seen as a weighted multi-label classification.


\section{Experiments}


\subsection{Dataset}

We perform our experiments on the \emph{Beatles} dataset as it provides the highest confidence regarding the ground truth annotations \cite{harte2010towards}. This dataset is composed by 180 songs annotated by hand. For each song, we compute the CQT by using a window size of 4096 samples and a hop size of 2048. The transform is mapped to a scale of 3 bins per semi-tone over 6 octaves ranging from C1 to C7. We augment the available data by performing all transpositions from -6 to +6 semi-tones and modifying the labels accordingly. Finally, to evaluate our models, we split the data into a training (60\%), validation (20\%) and test (20\%) sets.

\subsection{Models}

We use the same CNN model for all test configurations, but change the size of the last layer to fit the size of the selected chord alphabet. We apply a batch normalization and a Gaussian noise addition on the inputs layer. The architecture of the CNN consists of three convolutional layers followed by two fully-connected layers. The architecture is very similar to the first CNN that has been proposed for the ACE task \cite{humphrey2012rethinking}. However, we add dropout between each convolution layer to prevent over-fitting. 

For training, we use the ADAM optimizer with a learning rate of $2.10^{-5}$ for a total of 1000 epochs. We reduce the learning rate if the validation loss has not improved during 50 iterations. Early stopping is applied if the validation loss has not improved during 200 iterations and  we keep the model with the best validation accuracy. For each configuration, we perform a 5-cross validation by repeating a random split of the dataset.



\section{Results and discussion}

%

The aim of this paper is 
not to obtain the best classification scores (which would involve pre- or post-filtering methods)
but to study the impact on the classification results of different musical relationships (as detailed in the previous section). 
Therefore, we ran 9 instances of the CNN model corresponding to all combinations of the 3 alphabets $A_0$, $A_1$, $A_2$ and 3 distances $D_0$, $D_1$, $D_2$ to compare their results from both a \textit{quantitative} and \textit{qualitative} point of view.
We analyzed the results using the \emph{mireval} library \cite{raffel2014mir_eval} to compute classification scores, 
and a Python \textit{ACE Analyzer} that we developed to reveal the musical meaning of classification errors and, therefore, understand their qualities.


\subsection{Quantitative analysis: MIREX evaluation}

Regarding the MIREX evaluation, the efficiency of ACE models is assessed through classification scores over different alphabets \cite{raffel2014mir_eval}.
The MIREX alphabets for evaluation have a gradation of complexity from Major/Minor to Tetrads. 
In our case, for the evaluation on a specific alphabet, we apply a reduction from our training alphabet $A_i$ to the MIREX evaluation alphabet. Here, we evaluate on three alphabet : Major/Minor, Sevenths, and Tetrads. These alphabets correspond roughly to our three alphabets (Major/Minor $\sim$ $A_0$, Sevenths $\sim$ $A_1$, Tetrads $\sim$ $A_2$).

\subsubsection{MIREX Major/minor}
 
\begin{figure}
\centering
\includegraphics[width=0.45\textwidth]{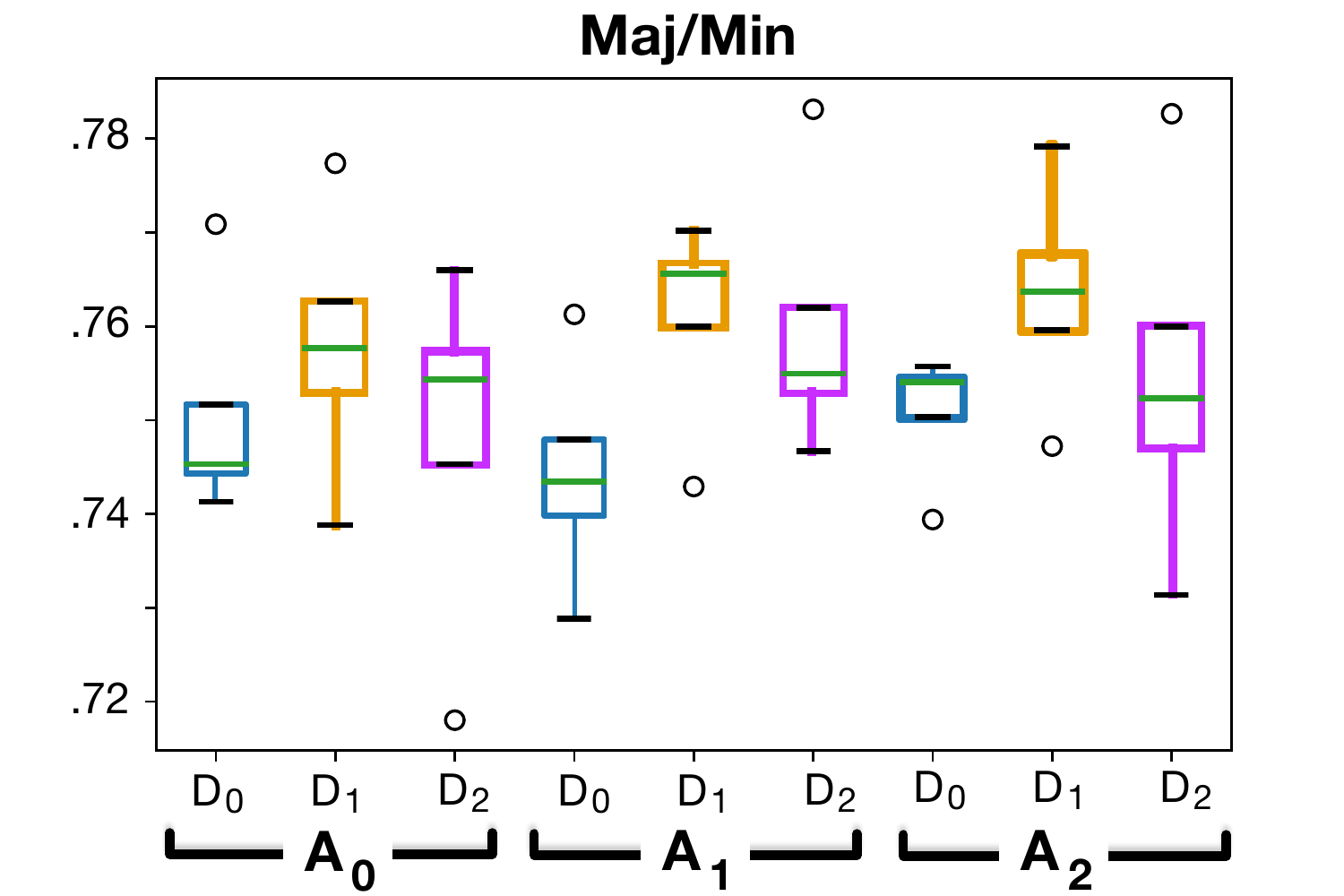}
\caption{\label{fig:majmin} Results of the 5-folds: evaluation on MIREX Maj/Min ($\sim$ reduction on $A_0$). }
\end{figure}

Figure~\ref{fig:majmin} depicts the average classification scores over all frames of our test dataset for different distances and alphabets. We can see that the introduction of the $D_1$ or $D_2$ distance improves the classification compared to $D_0$. With these distances, and even without pre- or post-filtering, we obtain classification scores that are superior to that of similar works (75.9\% for CNN with post-filtering but an extended dataset in \cite{humphrey2015four} versus 76.3\% for $A_2-D_1$).
Second, the impact of working first on large alphabets ($A_1$ and $A_2$), and then reducing on $A_0$ for the test is negligible on Maj/Min (only from a quantitative point of view, see \ref{subsec:qual}).


\subsubsection{MIREX Sevenths}
With more complex alphabets, the classification score is lower than for MIREX Maj/Min. This result is not surprising since we observe this behavior on all ACE systems. 
Moreover, the models give similar results and we can not observe a particular trend between the alphabet reductions or the different distances. 
The same result is observed for the evaluation with MIREX tetrads ($\sim$ reduction on $A_2$).
Nonetheless, the MIREX evaluation uses a binary score to compare chords. Because of this approach, the qualities of the classification errors cannot be evaluated. 


\begin{figure}
\centering
\includegraphics[width=0.45\textwidth]{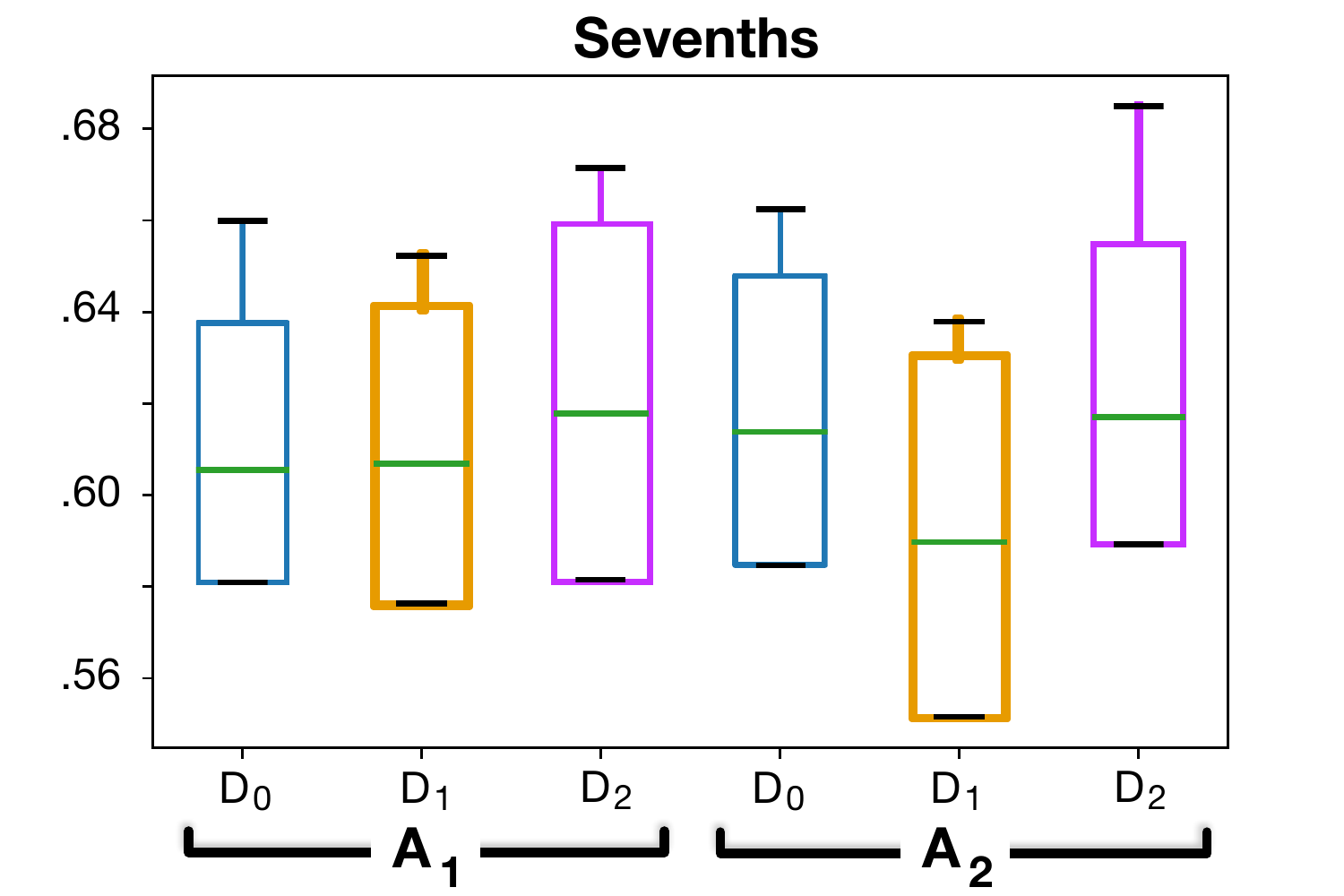}
\caption{\label{fig:sev} Results of the 5-folds: evaluation on MIREX Sevenths ($\sim$ reduction on $A_1$).}
\end{figure}

\subsection{Qualitative analysis: understanding the errors}\label{subsec:qual}

In this section, we propose to analyze ACE results from a qualitative point of view.
The aim here is not to introduce new alphabets or distances in the models, but to introduce a new type of evaluation of the results. Our goal is twofold: to understand what causes the errors in the first place, and to distinguish ``weak'' from ``strong'' errors with a \textit{functional} approach.

In tonal music, the \textit{harmonic functions} qualify the roles and the tonal significances of chords, 
and the possible equivalences between them within a sequence \cite{rehding2003hugo,schoenberg1969structural}.
Therefore, we developed an \textit{ACE Analyzer} including two modules discovering some formal musical relationships between the target chords and the chords predicted by ACE models. 
Both modules are generic and independent of the classification model, and are available online.\footnote{\url{http://repmus.ircam.fr/dyci2/ace_analyzer}}

\subsubsection{Substitution rules}

The first module detects the errors corresponding to hierarchical relationships or usual \textit{chord substitutions} rules: using a chord in place of another in a chord progression (usually substituted chords have two pitches in common with the triad that they are replacing).

%
Table~\ref{tab:sub2} presents: 
\textit{Tot.}, the total fraction of errors that can be explained by the whole set of substitution rules we implemented,
and \textit{$\subset$ \textit{Maj}} and \textit{$\subset$ \textit{min}}, 
the errors included in the correct triad 
(\textit{e.g.} \textit{C:maj} instead of \textit{C:maj7}, \textit{C:min7} instead of \textit{C:min}).
Table~\ref{tab:sub1} presents the percentages of errors corresponding to 
widely used substitution rules:
\textit{rel. m} and \textit{rel. M},
relative minor and major; 
\textit{T subs. 2},
tonic substitution different from \textit{rel. m} or \textit{rel. M} (\textit{e.g.} \textit{E:min7} instead or \textit{C:maj7}),
and the percentages of errors 
\textit{m$\rightarrow$M} and \textit{M$\rightarrow$m} (same root but major instead of minor or conversely). 
The tables only show the categories representing more than $1\%$ of the total number of errors, 
but other substitutions (that are not discussed here) were analyzed: tritone substitution, substitute dominant, and equivalence of  \textit{dim7} chords modulo inversions.

\begin{table}
\small
 \begin{center}
\begin{tabular}{l|c||c|c}
  \hline
  Model & \textit{Tot.} &  $\subset$ \textit{Maj} & $\subset$ \textit{min}  \\
\hline
$A_0$-$D_0$ & 34.93 & \_ & \_ \\
$A_0$-$D_1$ & 36.12 & \_ & \_ \\
$A_0$-$D_2$ & 35.37 & \_ & \_ \\
\hline
$A_1$-$D_0$ & 52.40 & 23.82 & 4.37  \\
$A_1$-$D_1$ & 57.67 & 28.31 & 5.37  \\
$A_1$-$D_2$ & 55.17 & 25.70 & 4.21  \\
\hline
$A_2$-$D_0$ & 55.28 & 26.51 & 4.29  \\
$A_2$-$D_1$ & 60.47 & 31.61 & 6.16  \\
$A_2$-$D_2$ & 55.45 & 25.74 & 4.78  \\
\hline

\end{tabular}
\end{center}
 \caption{Left: total percentage of errors corresponding to inclusions or chords substitutions rules, right: percentage of errors with inclusion in the correct triad (\% of the total number of errors).}
 \label{tab:sub2}
\end{table}

\begin{table}
\small
 \begin{center}
\begin{tabular}{l|c|c|c||c|c}
  \hline
  Model & \textit{rel. M} & \textit{rel. m} & \textit{T subs. 2} & \textit{m}$\rightarrow$\textit{M} & \textit{M}$\rightarrow$\textit{m}\\  
  \hline
  $A_0$-$D_0$ & 4.19 & 5.15 & 2.37 & 7.26 & 12.9 \\
  $A_0$-$D_1$ & 4.40 & 5.20 & 2.47 & 7.66 & 13.4 \\
  $A_0$-$D_2$ & 5.13 & 4.87 & 2.26 & 8.89 & 10.89 \\
  \hline
  $A_1$-$D_0$ & 2.63 & 3.93 & 1.53 & 4.46 & 8.83 \\
  $A_1$-$D_1$ & 3.05 & 3.36 & 1.58 & 5.53 & 7.52 \\
  $A_1$-$D_2$ & 3.02 & 4.00 & 1.62 & 5.84 & 8.07 \\
  \hline
  $A_2$-$D_0$ & 2.54 & 4.15 & 1.51 & 4.96 & 8.54 \\
  $A_2$-$D_1$ & 2.79 & 2.97 & 1.54 & 5.29 & 7.46 \\
  $A_2$-$D_2$ & 3.11 & 4.26 & 1.63 & 5.34 & 7.59 \\
  \hline
\end{tabular}
\end{center}
 \caption{Left: percentage of errors corresponding to usual chords substitutions rules, right: percentage of errors ``major instead of minor'' or inversely (\% of the total number of errors).}
 \label{tab:sub1}
\end{table}

First, \textit{Tot.} in Table~\ref{tab:sub2} shows that a huge fraction of errors can be explained by usual substitution rules. 
This percentage can reach 60.47\%, which means that numerous classification errors nevertheless give useful indications since they mistake a chord for another chord with an equivalent function. 
For instance, Table~\ref{tab:sub1} shows that a significant amount of errors (up to 10\%) are relative major / minor substitutions.
Besides, for the three distances, the percentage in \textit{Tot.} (Table~\ref{tab:sub2}) increases with the size of the alphabet: larger alphabets seem to imply weaker errors (higher amount of equivalent harmonic functions).

We can also note that numerous errors (between 28.19\% and 37.77\%) correspond to inclusions in major or minor chords (\textit{$\subset$ \textit{Maj}} and \textit{$\subset$ \textit{min}}, Table~\ref{tab:sub2}) for $A_1$ and $A_2$. 
In the framework of the discussion about \textit{recognition} and \textit{transcription} mentioned in introduction, this result questions the relevance of considering exhaustive extensions when the goal is to extract and formalize an underlying harmony.  

Finally, for $A_0$, $A_1$, and $A_2$, using $D_1$ instead of $D_0$ increases the fraction of errors attributed to categories in the left part of Table~\ref{tab:sub1} (and in almost all the configurations when using $D_2$).
This shows a qualitative improvement since all these operations are  considered as valid chord substitutions. 
On the other hand, the impact on the (quite high) percentages in the right part of Table~\ref{tab:sub1} is not clear. We can assume that temporal smoothing can be one of the keys to handle the errors \textit{m}$\rightarrow$\textit{M} and \textit{M}$\rightarrow$\textit{m}.
%

\subsubsection{Harmonic degrees}

The second module of our \textit{ACE Analyzer} focuses on \textit{harmonic degrees}. 
First, by using the annotations of key in the dataset in addition to that of chords, this module determines the roman numerals characterizing the harmonic degrees of the predicted chord and of the target chord (\textit{e.g.} in C, if a chord is an extension of \textit{C}, \textit{I}; if it is an extension of \textit{D:min}, \textit{ii}; etc.)
when it is possible
(\textit{e.g.} in C, if a chord is an extension of \textit{C\#} it does not correspond to any degree).
Then, it counts the errors corresponding to substitutions of harmonic degrees when it is possible (\textit{e.g.} in C, \textit{A:min} instead of \textit{C} corresponds to \textit{I$\sim$vi}).
%
%
This section shows an analysis of the results using this second module.
First, it determines if the target chord is diatonic (\textit{i.e.} belongs to the harmony of the key), as presented in Table~\ref{tab:deg1}.
If this is the case, the notion of incorrect degree for the predicted chord is relevant and the percentages of errors corresponding to substitutions of degrees is computed (Table~\ref{tab:deg2}).

\begin{table}
\small
 \begin{center}
\begin{tabular}{l|c||c}
  \hline
  Model & \textit{Non-diat. targ.} & \textit{Non-diat. pred.}\\
  \hline
  $A_0$-$D_0$ & 37.96 & 28.41 \\
  $A_0$-$D_1$ & 44.39 & 15.82\\
  $A_0$-$D_2$ & 45.87 & 17.60 \\
  \hline
  $A_1$-$D_0$ & 38.05 & 21.26 \\
  $A_1$-$D_1$ & 37.94 & 20.63 \\
  $A_1$-$D_2$ & 38.77 & 20.23 \\
  \hline
  $A_2$-$D_0$ & 37.13 & 30.01 \\
  $A_2$-$D_1$ & 36.99 & 28.41 \\
  $A_2$-$D_2$ & 37.96 & 28.24 \\
  \hline
\end{tabular}
\end{center}
 \caption{Errors occurring when the target is non-diatonic (\% of the total number of errors), non-diatonic prediction errors (\% of the subset of errors on diatonic targets).}
 \label{tab:deg1}
\end{table}

\begin{table}
\small
 \begin{center}
\begin{tabular}{l|c|c|c||c|c|c}
  \hline
  Model &\textit{I$\sim$IV}&\textit{I$\sim$V}&\textit{IV$\sim$V}&\textit{I$\sim$vi}&\textit{IV$\sim$ii}&\textit{I$\sim$iii}\\
  \hline
  $A_0$-$D_0$ & 17.41 & 14.04 & 4.54 & 4.22 & 5.41 & 2.13 \\
  $A_0$-$D_1$ & 17.02 & 13.67 & 3.33 & 4.08 & 6.51 & 3.49\\
  $A_0$-$D_2$ & 16.16 & 13.60 & 3.08 & 5.65 & 6.25 & 3.66\\
  \hline
  $A_1$-$D_0$ & 17.53 & 13.72 & 3.67 & 5.25 & 4.65 & 3.50\\
  $A_1$-$D_1$ & 15.88 & 13.82 & 3.48 & 4.95 & 6.26 & 3.46\\
  $A_1$-$D_2$ & 16.73 & 13.45 & 3.36 & 4.70 & 5.75 & 2.97	\\
  \hline
  $A_2$-$D_0$ & 16.90 & 13.51 & 3.68 & 4.45 & 5.06 & 3.32\\
  $A_2$-$D_1$ & 16.81 & 13.60 & 3.85 & 4.57 & 5.37 & 3.59 \\
  $A_2$-$D_2$ & 16.78 & 12.96 & 3.84 & 5.19 & 7.01 & 3.45 \\
  \hline
\end{tabular}
\end{center}
 \caption{Errors ($> 2\%$) corresponding to degrees substitutions (\% of the subset of errors on diatonic targets).}
 \label{tab:deg2}
\end{table}

A first interesting fact presented in Table~\ref{tab:deg1} is that 36.99\% to 45.87\% of the errors occur when the target chord is non-diatonic.
It also shows, for the three alphabets, that using $D_1$ or $D_2$ instead of $D_0$ makes the fraction of non-diatonic errors decrease (Table~\ref{tab:deg1}, particularly $A_0$), which means that the errors are more likely to stay in the correct key. 
Surprisingly, high percentages of errors are associated to errors \textit{I$\sim$V} (up to 14.04\%), \textit{I$\sim$IV} (up to 17.41\%), or \textit{IV$\sim$V} (up to 4.54\%) in Table~\ref{tab:deg2}. These errors are not usual substitutions, and \textit{IV$\sim$V} and \textit{I$\sim$IV} have respectively 0 and 1 pitch in common. 
In most of the cases, these percentages tend to decrease on alphabets $A_1$ or $A_2$ and when using musical distances (particularly $D_2$). Conversely, it increases the amount of errors in the right part of Table~\ref{tab:deg2} containing usual substitutions: 
once again we observe that
the more precise the musical representations are, the more the harmonic functions tend to be correct.


\section{Conclusion}






We presented a novel approach taking advantage of musical prior knowledge underlying the labeling alphabets into ACE statistical models. To this end, we applied reductions on different chord alphabets
and we 
used different distances
to train the same type of model. Then, we conducted a quantitative and qualitative analysis of the classification results.

First, we conclude that training the model using distances reflecting the relationships between chords improves the results both quantitatively (classification scores) and qualitatively (in terms of harmonic functions). 
%
%
Second, it appears that
working first on large alphabets and reducing the chords during the test phase 
does not significantly improve the classification scores 
but provides a qualitative improvement in the type of errors.
%
%
%
Finally, ACE could be improved by moving away from its binary classification paradigm.
Indeed, MIREX evaluations focus on the nature of chords but a large amount of errors can be explained by
inclusions or usual substitution rules.
Our evaluation method therefore
provides an interesting notion of \textit{musical quality} of the errors, and encourages to adopt a functional approach or even to introduce a notion of equivalence classes.
It could be adapted to the ACE problem downstream and upstream: in the classification processes as well as in the methodology for labeling the datasets.

\section{Acknowledgments}
The authors would like to thank the master's students who contributed to the implementation: Alexis Font, Gr{\'e}goire Locqueville, Octave Roulleau-Thery, and T{\'e}o Sanchez.
This work was supported by 
the DYCI2 project ANR-14-CE2 4-0002-01 funded by the French National Research Agency (ANR),  
the MAKIMOno project 17-CE38-0015-01 funded by the French ANR and the Canadian Natural Sciences and Engineering Reserch Council (STPG 507004-17), the ACTOR Partnership funded by the Canadian Social Sciences and Humanities Research Council (895-2018-1023).


%


\bibliography{main}

\begin{thebibliography}{10}

\bibitem{boulanger2013audio}
N.~Boulanger-Lewandowski, Y.~Bengio, and P.~Vincent.
\newblock Audio chord recognition with recurrent neural networks.
\newblock In {\em International Symposium on Music Information Retrieval},
  pages 335--340, 2013.

\bibitem{cannam2015mirex}
C.~Cannam, E.~Benetos, M.~Mauch, M.~E.~P. Davies, S.~Dixon, C.~Landone,
  K.~Noland, and D.~Stowell.
\newblock Mirex 2015: Vamp plugins from the centre for digital music.
\newblock {\em In Proceedings of the Music Information Retrieval Evaluation
  eXchange (MIREX)}, 2015.

\bibitem{cho2014improved}
T.~Cho.
\newblock {\em Improved techniques for automatic chord recognition from music
  audio signals}.
\newblock PhD thesis, New York University, 2014.

\bibitem{cho2010exploring}
T.~Cho, R.~J. Weiss, and J.~P. Bello.
\newblock Exploring common variations in state of the art chord recognition
  systems.
\newblock In {\em Proceedings of the Sound and Music Computing Conference
  (SMC)}, pages 1--8, 2010.

\bibitem{cohn1997neo}
R.~Cohn.
\newblock Neo-riemannian operations, parsimonious trichords, and their"
  tonnetz" representations.
\newblock {\em Journal of Music Theory}, 41(1):1--66, 1997.

\bibitem{harte2010towards}
C.~Harte.
\newblock {\em Towards automatic extraction of harmony information from music
  signals}.
\newblock PhD thesis, 2010.

\bibitem{harte2005symbolic}
C.~Harte, M.~B. Sandler, S.~A. Abdallah, and E.~G{\'o}mez.
\newblock Symbolic representation of musical chords: A proposed syntax for text
  annotations.
\newblock In {\em International Symposium on Music Information Retrieval},
  volume~5, pages 66--71, 2005.

\bibitem{huang2016chordripple}
C-Z.~A. Huang, D.~Duvenaud, and K.~Z. Gajos.
\newblock Chordripple: Recommending chords to help novice composers go beyond
  the ordinary.
\newblock In {\em Proceedings of the 21st International Conference on
  Intelligent User Interfaces}, pages 241--250. ACM, 2016.

\bibitem{humphrey2012rethinking}
E.~J. Humphrey and J.~P. Bello.
\newblock Rethinking automatic chord recognition with convolutional neural
  networks.
\newblock In {\em Machine Learning and Applications (ICMLA), 2012 11th
  International Conference on}, volume~2, pages 357--362. IEEE, 2012.

\bibitem{humphrey2015four}
E.~J. Humphrey and J.~P. Bello.
\newblock Four timely insights on automatic chord estimation.
\newblock In {\em International Symposium on Music Information Retrieval},
  pages 673--679, 2015.

\bibitem{humphrey2012learning}
E.~J. Humphrey, T.~Cho, and J.~P. Bello.
\newblock Learning a robust tonnetz-space transform for automatic chord
  recognition.
\newblock In {\em Acoustics, Speech and Signal Processing (ICASSP), 2012 IEEE
  International Conference on}, pages 453--456. IEEE, 2012.

\bibitem{jiang2017extended}
J.~Jiang, W.~Li, and Y.~Wu.
\newblock Extended abstract for mirex 2017 submission: Chord recognition using
  random forest model.
\newblock {\em MIREX evaluation results}, 2017.

\bibitem{korzeniowski2016feature}
F.~Korzeniowski and G.~Widmer.
\newblock Feature learning for chord recognition: The deep chroma extractor.
\newblock {\em arXiv preprint arXiv:1612.05065}, 2016.

\bibitem{korzeniowski2016fully}
F.~Korzeniowski and G.~Widmer.
\newblock A fully convolutional deep auditory model for musical chord
  recognition.
\newblock In {\em Machine Learning for Signal Processing (MLSP), 2016 IEEE 26th
  International Workshop on}, pages 1--6. IEEE, 2016.

\bibitem{lafferty2001conditional}
J.~Lafferty, A.~McCallum, and F.~C.~N. Pereira.
\newblock Conditional random fields: Probabilistic models for segmenting and
  labeling sequence data.
\newblock 2001.

\bibitem{lee2006automatic}
K.~Lee.
\newblock Automatic chord recognition from audio using enhanced pitch class
  profile.
\newblock In {\em International Computer Music Conference}, 2006.

\bibitem{lou1995implementing}
H-L. Lou.
\newblock Implementing the viterbi algorithm.
\newblock {\em IEEE Signal Processing Magazine}, 12(5):42--52, 1995.

\bibitem{madjiheurem2016chord2vec}
S.~Madjiheurem, L.~Qu, and C.~Walder.
\newblock Chord2vec: Learning musical chord embeddings.
\newblock In {\em Proceedings of the Constructive Machine Learning Workshop at
  30th Conference on Neural Information Processing Systems (NIPS’2016),
  Barcelona, Spain}, 2016.

\bibitem{mcfee2017structured}
B.~McFee and J.~P. Bello.
\newblock Structured training for large-vocabulary chord recognition.
\newblock In {\em Proceedings of the 18th International Society for Music
  Information Retrieval Conference (ISMIR’2017). ISMIR}, 2017.

\bibitem{mcvicar2014automatic}
M.~McVicar, R.~Santos-Rodr{\'\i}guez, Y.~Ni, and T.~De Bie.
\newblock Automatic chord estimation from audio: A review of the state of the
  art.
\newblock {\em IEEE/ACM Transactions on Audio, Speech and Language Processing
  (TASLP)}, 22(2):556--575, 2014.

\bibitem{oudre2009template}
L.~Oudre, Y.~Grenier, and C.~F{\'e}votte.
\newblock Template-based chord recognition: Influence of the chord types.
\newblock In {\em International Symposium on Music Information Retrieval},
  pages 153--158, 2009.

\bibitem{raffel2014mir_eval}
C.~Raffel, B.~McFee, E.~J. Humphrey, J.~Salamon, O.~Nieto, D.~Liang, D.~P.~W.
  Ellis, and C.~C. Raffel.
\newblock mir\_eval: A transparent implementation of common mir metrics.
\newblock In {\em Proceedings of the 15th International Society for Music
  Information Retrieval Conference, ISMIR}, 2014.

\bibitem{rehding2003hugo}
A.~Rehding.
\newblock {\em Hugo Riemann and the birth of modern musical thought},
  volume~11.
\newblock Cambridge University Press, 2003.

\bibitem{schoenberg1969structural}
A.~Schoenberg and L.~Stein.
\newblock {\em Structural functions of harmony}.
\newblock Number 478. WW Norton \& Company, 1969.

\bibitem{wu2017mirex}
Y.~Wu, X.~Feng, and W.~Li.
\newblock Mirex 2017 submission: Automatic audio chord recognition with
  miditrained deep feature and blstm-crf sequence decoding model.
\newblock {\em MIREX evaluation results}, 2017.

\bibitem{zhou2015chord}
X.~Zhou and A.~Lerch.
\newblock Chord detection using deep learning.
\newblock In {\em Proceedings of the 16th International Symposium on Music
  Information Retrieval Conference}, volume~53, 2015.

\end{thebibliography}

%
%
%
%

\end{document}